\documentclass[letterpaper, 11 pt]{article}                                                          

\usepackage{graphicx}                                                          
\usepackage{amsmath,amssymb}                                                         

\date{}

\title{\Huge
Static Output Feedback Control for Nonlinear Systems subject to Parametric and Nonlinear Uncertainties}

\author{Masoud Abbaszadeh and Horacio J. Marquez
\thanks{M. Abbaszadeh, {\tt\small e-mail: masoud@ualberta.net}}%
\thanks{H. J. Marquez is with the Department of Electrical and Computer Engineering,
        University of Alberta, Edmonton, Alberta, Canada, T6G 2V4,
        {\tt\small e-mail: marquez@ece.ualberta.ca}}%
}

\begin{document}

\maketitle \thispagestyle{empty} \pagestyle{empty}

\begin{abstract}
This work addresses the design of static output feedback
control of discrete-time nonlinear systems satisfying a local Lipschitz continuity
condition with time-varying uncertainties. The controller has also a guaranteed disturbance
attenuation level ($H_{\infty}$ performance). Thanks to the
linearity of the proposed LMIs in both the admissible Lipschitz
constant of the system and the disturbance attenuation level, they
can be simultaneously optimized through convex multiobjective
optimization. The optimization over Lipschitz constant adds an extra
important and new feature to the controller, robustness against
nonlinear uncertainty. The resulting controller is robust against
both nonlinear additive uncertainty and time-varying parametric
uncertainties. Explicit norm-wise and element-wise bounds on the
tolerable nonlinear uncertainty are derived.
\end{abstract}


\section{Introduction}
The static output feedback (SOF) stabilization problem
is known to be a challenging task and in spite of receiving great
attention it is still one of the most important open questions in
the control theory. Over the past decays most advances in the
control theory have been limited to the state feedback control and
dynamic output feedback control. However, state feedback techniques,
require either the measurement of every system state some of which
expensive or even impossible to be measured or using the observer
based controllers which makes the implementation task expensive and
hard. On the other hand, dynamic output feedback designs, result in
high order controllers which may not be desirable in many industrial
applications again due to implementation difficulties. Controllers
using static output feedback are less expensive to implement and are
more reliable. Therefore, many researchers have tried to
characterize the problem of finding a stabilizing SOF controller
\cite{el1997cone, cao1998static, toker1995np, kau2007robust, geromel1998static,
qiu2010fuzzy, henrion2006convergent, iwasaki1994linear, li2014heuristic,
dong2013robust, rubio2013static, aouaouda2014robust}.
A comprehensive survey on static output feedback is given in
\cite{Syrmos}.

Despite abundant literature, due to the non-convexity of the SOF
formulation, the problem is still open both analytically and
numerically. While the necessary and sufficient conditions of the
existence of a stabilizing SOF controller based on the original
non-convex formulation are not numerically tractable, the existing
convex formulations often lead to restrictive sufficient conditions.
The proposed results include methods based on structural pole
assignment \cite{Yang, Franze}, Riccati-based approaches \cite{Neto,
Kucera} and optimization formulation (min-max problem)
\cite{Geromel}. The original non-convex problem can be directly
formulated using bilinear matrix inequalities (BMIs). However, the
numerical solution of BMI problems has been shown to be NP-hard
\cite{Toker}. Due to recent advances in linear matrix inequalities
both theoretically and numerically, several works have been recently
addressed in the literature attempting to cast the SOF problem into
the convex LMI framework. Available methods include using ILMIs
(iterative LMIs) \cite{Fujimori}, \cite{Cao} where there is no
guarantee for the convergence of iterations or imposing
nonsingularity conditions over the state space realization matrices
\cite{Garcia} or a submatrix of ``$A$'' \cite{Bara1}.

Alternatively, is has been shown that in some cases the BMI problem
can be converted into a semidefinite cone programming problem (SDP)
\cite{Mesbahi}. The advantage gained through this conversion is that
reliable algorithms exist to solve SDP problems numerically. In this
work, we propose a novel method for robust static output feedback
stabilization of a class of discrete-time nonlinear uncertain
systems. The method proposed in this paper is non-iterative and not
only provides a less restrictive solution but also extends the
results to a more general class of systems where there are
parametric uncertainties and Lipschitz nonlinearity in the model. In
addition the proposed controller has a guaranteed disturbance
attenuation level ($H_{\infty}$ performance). Our goal is to develop
linear matrix inequalities in which the Lipschitz constant is one
the LMI variables in order to achieve the maximum admissible
Lipschitz constant through convex optimization. This optimization
adds an important extra feature to the SOF controller making it
robust against nonlinear uncertainties. Explicit bounds on the
tolerable nonlinear uncertainty are derived through norm-wise and
element-wise robustness analysis. Actually, thanks to the linearity
of the proposed LMIs in both the admissible Lipschitz constant of
the system and the disturbance attenuation level, they can be
simultaneously optimized through convex multiobjective optimization.
In the next stage, we show that the proposed solution can be
modified as an SDP problem. In fact original BMI problem is
converted into an SDP problem. The paper is organized as follows. In
section II, the problem statement and some preliminaries are
mentioned. In Section III, we propose a new method for robust
$H_{\infty}$ SOF controller design for nonlinear uncertain systems.
Section IV, is devoted to robustness analysis in which explicit
bounds on the tolerable nonlinear uncertainty are derived. Section
V, contains an illustrative example showing the high performance of
our proposed method.

\section{Preliminaries and Problem Statement}
Consider the following class of nonlinear discrete-time uncertain
systems:
\begin{align}
\left({\sum}_{1}\right): \ x(k+1)&=(A+\Delta
A(k))x(k)+\Phi(x,u)\notag\\ & \ \ \ +B_{1}u(k)+B_{2}w(k) \label{sigma3_1}\\
y(k)&=(C+\Delta C(k))x(k)+Dw(k) \label{output1}\\
z(k)&=Hx(k)
\end{align}
where $x\in {\mathbb R} ^{n} ,u\in {\mathbb R} ^{m} ,y\in {\mathbb
R} ^{p} $ and $\Phi(x,u)$ contains nonlinearities of second order or
higher. We assume that the system $({\sum}_{1})$ is locally
Lipschitz with respect to $x$ in a region $\mathcal{D}$ containing
the origin, uniformly in $u$, i.e.:
\begin{align}
&\|\Phi(0,u^{*})\|=0 \label{Lip1}\\
&\|\Phi(x_{1},u^{*})-\Phi(x_{2},u^{*})\|\leqslant
\gamma\|x_{1}-x_{2}\| \hspace{4mm}\forall \, x(k)\in
\mathcal{D}\label{Lip2}
\end{align}
where $\|.\|$ is the induced 2-norm, $u^{*}$ is any admissible
control signal and $\gamma>0$ is called the Lipschitz constant. If
the nonlinear function $\Phi$ satisfies the Lipschitz continuity
condition globally in $\mathbb{R}^{n}$, then the results will be
valid globally. We assume that $rank(B_{1})=m<n$ and $rank(C)=p<n$.
$w(k)\in\ell_{2}[0,\infty)$ is an unknown exogenous disturbance and
$\Delta A(k)$ and $\Delta C(k)$ are unknown matrices representing
time-varying parameter uncertainties, and are assumed to be of the
form
\begin{align}
\left[
  \begin{array}{c}
    \Delta A(k) \\
    \Delta C(k) \\
  \end{array}
\right]= \left[
  \begin{array}{c}
    M_{1} \\
    M_{2} \\
  \end{array}
\right]F(k)N \label{uncer1}
\end{align}
where $M_{1}$, $M_{2}$ and $N$ are known real constant matrices and
$F(k)$ is an unknown real-valued time-varying matrix satisfying
\begin{equation}
\forall k, \ \ F^{T}(k)F(k)\leq I.\notag
\end{equation}
It is also worth noting that the structure of parameter
uncertainties in (\ref{uncer1}) has been widely used in the problems
of robust control and robust filtering for both continuous-time and
discrete-time systems and can capture the uncertainty in several
practical situations \cite{Khargonekar}. Suppose $z(k)=Hx(k)$ stands
for the controlled output for state where $H$ is a known matrix. Our
purpose is to design the static gain $K$ such that the static output
feedback control law $u(k)=Ky(k)$ robustly asymptotically stabilizes
the system with maximum admissible Lipschitz constant in the
presence of uncertainty and the following specified $H_{\infty}$
norm upper bound is simultaneously guaranteed.
\begin{equation}
\|z\|\leq\mu\|w\|.
\end{equation}

In the following, we mention some useful lemmas that will be used
later in the proof of our results. \\

\emph{\textbf{Lemma 1. \cite{deSouza2}} For any
$x,y\in\mathbb{R}^{n}$ and any positive definite matrix
$P\in\mathbb{R}^{n\times{n}}$, we have}
\begin{equation}
2x^{T}y\leq x^{T}Px+y^{T}P^{-1}y.\notag
\end{equation}

\emph{\textbf{Lemma 2. \cite{deSouza2}} Let $A,D, E, F$ and P be
real matrices of appropriate dimensions with $P>0$ and $F$
satisfying $F^{T}F\leq I$. Then for any scalar $\epsilon>0$
satisfying $P^{-1}-\epsilon^{-1}DD^{T}>0$, we have}
\begin{equation}
\begin{split}
(A+DFE)^{T}P(A+DFE)\leq
A^{T}(P^{-1}-\epsilon^{-1}DD^{T})^{-1}A\\+\epsilon E^{T}E.\notag
\end{split}
\end{equation}

\subsection{Notation}
The following matrix notation will be used throughout the paper. For
matrices $A=[a_{ij}]_{m\times n}$ and $B=[b_{ij}]_{m\times n}$,
$A\preceq B$ means $a_{i,j}\leq b_{ij} \ \forall \ 1\leq i \leq m
,1\leq j\leq n$. For square A, $diag (A)$ is a vector containing the
elements on the main diagonal and $diag(x)$ where $x$ is a vector is
a diagonal matrix with the elements of $x$ on the main diagonal.
$|A|$ is the element-wise absolute value of $A$, i.e.
$[|a_{ij}|]_{n}$. $A\circ B$ stands for the element-wise product
\emph{(Hadamard product)} of $A$ and $B$. $A\otimes B$ denotes the
\emph{Kronecker (tensor) product} of $A=[a_{ij}]_{m\times n}$ and
$B_{p\times q}$ as $A\otimes B=[a_{ij}B]_{mp\times nq}$.
$\textrm{\textbf{vec}}(A)$ is the vector obtained by stacking up the
columns of the matrix $A$.

\section{Static Output Feedback Stabilization}
In this section we propose a novel method for robust output feedback
stabilization for nonlinear discrete-time uncertain systems.
Consider the nonlinear uncertain system of class $({\sum}_{1})$. The
LMI approach for the static output feedback stabilization problem is
still unsolved due to its bilinear nature. Available methods include
using ILMIs (iterative LMIs) \cite{Fujimori}, \cite{Cao} where there
is no guarantee for the convergence of iterations or imposing
nonsingularity conditions over either $B_{1}$, $C$, $A$
\cite{Garcia} or a submatrix of $A$ \cite{Bara1}. The following
theorem not only provides a new solution for the problem but also
extends the result to the case where there exist time-varying
uncertainties in the pair $(A,C)$ and Lipschitz nonlinearity in the
state dynamics. As mentioned earlier, our goal is to design a robust
stabilizing controller with the $H_{\infty}$ performance $\|z\|\leq
\mu \|w\|$, for systems of class $\left({\sum}_{1}\right)$, using
static output feedback. \\
We first prove a lemma about robust asymptotic stability in the
presence of exogenous disturbance.\\

\emph{\textbf{Lemma 3.} Consider the following nonlinear uncertain
system
\begin{align}
\left({\sum}_{2}\right): \ x(k+1)&=(A+\Delta
A(k))x(k)\notag\\&\hspace{2cm}+\Phi(x,u)+Bw\label{sys5}\\
z(k)&=Hx(k).
\end{align}
This system is asymptotically stable with $\|z\|\leq \mu\|w \|$ and
maximum admissible Lipschitz constant $\gamma^{*}$, if there exist
scalars $\alpha>0$, $\epsilon_{1}>0$ and $\epsilon_{2}>0$ and a
matrix $P>0$
such that the following LMI optimization problem has a solution:}\\
\begin{equation}
\hspace{-6cm} \min (\alpha+\epsilon_{1}) \notag
\end{equation}
\hspace{.18cm} \emph{s.t.}
\begin{align}
&\left[
  \begin{array}{ccccc}
    \Lambda_{1} & I & A^{T}P & 0 & 0 \\
    \star & -\alpha I & 0 & 0 & 0 \\
    \star & \star & -\frac{1}{2}P & P & PM_{1} \\
    \star & \star & \star & P-2\epsilon_{1}I & 0 \\
    \star & \star & \star & \star & -\epsilon_{2} I \\
  \end{array}
\right]<0\label{LMI6}\\
&\left[
  \begin{array}{ccc}
    -\mu^{2}I & B^{T}P & B^{T}P\\
    \star & -\frac{1}{2}P & 0 \\
    \star & \star & -I
  \end{array}
\right]<0\label{LMI7}
\end{align}
\emph{where $\Lambda_{1}=H^{T}H-P+\epsilon_{2} N^{T}N$. Once the
problem is solved}
\begin{eqnarray}
\alpha^{*} &\triangleq& \min(\alpha)\notag
\\\epsilon_{1}^{*} &\triangleq& \min(\epsilon_{1})\notag
\\\gamma^{*} &\triangleq& \max(\gamma)=\frac{1}{\sqrt{\alpha^{*}(1+\epsilon_{1}^{*})}}\notag
\end{eqnarray}\\
\textbf{Proof:} Consider the Lyapunov function candidate:
\begin{eqnarray}
V=x^{T}Px.
\end{eqnarray}
We have
\begin{multline}
\Delta V_{k}=x^{T}(A+\Delta A)^{T}P(A+\Delta A)x\\+2x^{T}(A+\Delta
A)^{T}P\Phi+\Phi^{T}P\Phi-x^{T}Px+2w^{T}B^{T}P\Phi
\\+2w^{T}B^{T}P(A+\Delta A)x+w^{T}B^{T}PBw.
\label{deltav2}
\end{multline}
\begin{align}
\begin{split}
(A+\Delta A)^{T}P(A+\Delta A)&=(A+\Delta A)^{T}PP^{-1}P(A+\Delta A)\\
&=(PA+PM_{1}FN)^{T}P^{-1}(PA+PM_{1}FN).
\end{split}
\end{align}
Based on Lemma 1 and \eqref{Lip1}-\eqref{Lip2},
\begin{equation}
\begin{split}
2&x^{T}(A+\Delta A)^{T}P\Phi+\Phi^{T}P\Phi= 2x^{T}(A+\Delta
A)^{T}P\Phi\\&-\Phi^{T}W\Phi+\epsilon_{1}\Phi^{T}\Phi \leq
x^{T}(A+\Delta A)^{T}PW^{-1}P(A+\Delta
A)x\\&+\epsilon_{1}\gamma^{2}x^{T}x
\end{split}\notag
\end{equation}
where $W\triangleq\epsilon_{1}I-P$. It can be concluded from
\eqref{LMI6} that
$\frac{1}{2}P-(2\epsilon_{1}I-P)^{-1}P^{2}-\epsilon_{2}^{-1}
PM_{1}M_{1}^{T}P>0$ and $2\epsilon_{1}I-P>0$. Therefore, since
$P>0$, the condition $W>0$ is already included in \eqref{LMI6}.
Thus,
\begin{equation}
\begin{split}
x^{T}&(A+\Delta A)^{T}P(A+\Delta A)x+2x^{T}(A+\Delta A)^{T}P\Phi\\
&+\Phi^{T}P\Phi-x^{T}Px\leq
x^{T}[(PA+PM_{1}FN)^{T}\cdots\\&(P^{-1}+W^{-1})(PA+PM_{1}FN)-P+\epsilon_{1}\gamma^{2}I]x.
\label{ineq1}
\end{split}
\end{equation}
Using Lemma 1, it can be written
\begin{equation}
\begin{split}
2&w^{T}B^{T}P(A+\Delta A)x+2w^{T}B^{T}P\Phi+w^{T}B^{T}PBw\\&\leq
2w^{T}B^{T}P(A+\Delta A)x+w^{T}B^{T}PPBw+w^{T}B^{T}PBw\\
&+\Phi^{T}\Phi\leq 2w^{T}B^{T}P(A+\Delta
A)x+w^{T}B^{T}PPBw\\&+w^{T}B^{T}PBw+\gamma^{2}x^{T}x \leq
x^{T}(A+\Delta A)^{T}P(A+\Delta
A)x\\&+w^{T}[B^{T}PB+B^{T}PPB+B^{T}PB]w+\gamma^{2}x^{T}x\\
&= x^{T}[(A+\Delta A)^{T}PP^{-1}P(A+\Delta
A)+\gamma^{2}]x^{T}\\&+w^{T}[2B^{T}PP^{-1}PB+B^{T}PPB]w.\label{ineq2}
\end{split}
\end{equation}
Based on Lemma 1,
\begin{eqnarray}
2x^{T}Py&\leq& x^{T}Px+y^{T}Py\\
2x^{T}P^{-1}y&\leq& x^{T}P^{-1}x+y^{T}Py \label{ineq4}
\end{eqnarray}
from the \eqref{ineq4} and since $P<I$,
\begin{equation}
\begin{split}
&2w^{T}B^{T}P(A+\Delta A)x+2w^{T}B^{T}P\Phi+w^{T}B^{T}PBw\\&\leq
2w^{T}B^{T}P(A+\Delta
A)x+w^{T}B^{T}PPBw+w^{T}B^{T}PBw+\Phi^{T}\Phi\\
&\leq 2w^{T}B^{T}P(A+\Delta
A)x+w^{T}B^{T}PPBw+w^{T}B^{T}PBw+\gamma^{2}x^{T}x\\
&\leq x^{T}(A+\Delta A)^{T}PP(A+\Delta
A)x+w^{T}[B^{T}PP^{-1}PB+B^{T}PPB+B^{T}PB]w+\gamma^{2}x^{T}x\\
&\leq x^{T}(A+\Delta A)^{T}PP(A+\Delta
A)x+3w^{T}B^{T}PP^{-1}PBw+\gamma^{2}x^{T}x.\label{ineq2}
\end{split}
\end{equation}
Substituting \eqref{ineq1} and \eqref{ineq2} into \eqref{deltav2}
yields to
\begin{multline}
\Delta V_{k}\leq
x^{T}[(PA+PM_{1}FN)^{T}(2P^{-1}+W^{-1})\cdots\\(PA+PM_{1}FN)-P+(1+\epsilon_{1})\gamma^{2}I]x+3w^{T}B^{T}PP^{-1}PBw.\notag
\end{multline}
Now, define
\begin{equation}
J\triangleq
\sum_{k=0}^{\infty}\left[z(k)^{T}z(k)-\mu^{2}w(k)^{T}w(k)\right]\notag
\end{equation}
so,
\begin{equation}
J < \sum_{k=0}^{\infty}\left[z(k)^{T}z(k)-\mu^{2}w(k)^{T}w(k)+\Delta
V\right]\notag
\end{equation}
thus, a sufficient condition for $J\leq0$ is that
\begin{equation}
\begin{split}
x^{T}[(PA&+PM_{1}FN)^{T}(2P^{-1}+W^{-1})(PA\cdots\\&+PM_{1}FN)+H^{T}H-P+(1+\epsilon_{1})\gamma^{2}I]x\\
&+w^{T}[3B^{T}PP^{-1}PB-\mu^{2}I]w\leq 0 \label{ineq3}.
\end{split}
\end{equation}
Defining the new variable
\begin{equation}
\alpha\triangleq\frac{1}{(1+\epsilon_{1})\gamma^{2}}\Rightarrow
\gamma=\frac{1}{\sqrt{\alpha(1+\epsilon_{1})}}\label{alpha_def},
\end{equation}
maximization of $\gamma$ is equivalent to the simultaneous
minimization of $\alpha$ and $\epsilon_{1}$. Combining the two
objective functions, we minimize the scalarized linear objective
function $\alpha+\epsilon_{1}$. On the other hand,
\begin{equation}
\begin{split}
2P^{-1}&+(\epsilon_{1}I-P)^{-1}=P^{-1}+P^{-1}+(\epsilon_{1}I-P)^{-1}\\
&=P^{-1}+(\epsilon_{1}I-P)^{-1}[(\epsilon_{1}I-P)P^{-1}+I]\\
&=P^{-1}+\epsilon_{1}(\epsilon_{1}I-P)^{-1}P^{-1}\\
&=P^{-1}+(P-\epsilon_{1}^{-1}P^{2})^{-1}\\
&=(P-\epsilon_{1}^{-1}P^{2})^{-1}[(P-\epsilon_{1}^{-1}P^{2})P^{-1}+I]\\
&=(P-\epsilon_{1}^{-1}P^{2})^{-1}(2I-\epsilon_{1}^{-1}P)\\
&\Rightarrow\\
[2P^{-1}+&(\epsilon_{1}I-P)^{-1}]^{-1}=(2I-\epsilon_{1}^{-1}P)^{-1}(P-\epsilon_{1}^{-1}P^{2})\\
&=(2I-\epsilon_{1}^{-1}P)^{-1}P-(2\epsilon_{1}I-P)^{-1}P^{2}\\
&> \frac{1}{2}P-(2\epsilon_{1}I-P)^{-1}P^{2}\Rightarrow\\
2P^{-1}&+(\epsilon_{1}I-P)^{-1}<[\frac{1}{2}P-(2\epsilon_{1}I-P)^{-1}P^{2}]^{-1}
\end{split}\notag
\end{equation}
So, according to \eqref{ineq3}, a sufficient condition for $J\leq0$
is that
\begin{multline}
x^{T}\{(PA+PM_{1}FN)^{T}[\frac{1}{2}P-(2\epsilon_{1}I-P)^{-1}P^{2}]^{-1}(PA\cdots\\+PM_{1}FN)+H^{T}H-P+\alpha^{-1}I\}x
+w^{T}[2B^{T}PP^{-1}PB\cdots\\+B^{T}PPB-\mu^{2}I]w< 0.\notag
\end{multline}
According to Lemma 2,
\begin{eqnarray}
\begin{split}
x^{T}\{A^{T}P&[\frac{1}{2}P-(2\epsilon_{1}I-P)^{-1}P^{2}-\epsilon_{2}^{-1}
PM_{1}M_{1}^{T}P]^{-1}PA
\\&+\epsilon_{2} N^{T}N+H^{T}H-P+\alpha^{-1}I\}x\\&+w^{T}[2B^{T}PP^{-1}PB+B^{T}PPB-\mu^{2}I]w < 0
\end{split}\notag
\end{eqnarray}
which is by Schur complements equivalent to LMIs \eqref{LMI6} and
\eqref{LMI7}. $\blacksquare$\\

\textbf{Remark 1.} Lemma 3, provides a tool for robust stability
analysis of the aforementioned class of nonlinear systems.
Maximization of $\gamma$ guarantees the stability of the system for
any Lipschitz nonlinear function with Lipschitz constant less than
or equal $\gamma^{*}$. It is clear that if the stability of a
systems with a given fixed Lipschitz constant is to be analyzed, the
proposed LMI optimization problem will reduce to an LMI feasibility
problem and there will be no need to the change of variable
\eqref{alpha_def} anymore.\\

\textbf{Remark 2.} The proposed LMIs are linear in $\alpha$,
$\epsilon_{1}$ and $\zeta(=\mu^{2})$. Thus, either can be a fixed
constant or an optimization variable.\\

\subsection{Non-iterative Strict LMI solution}

The static output feedback problem is bilinear by its nature. On the
other hand, it has been shown that the solution of the bilinear
matrix inequalities is NP-hard. Here we propose a non-interactive
strict LMI solution to the problem which can be solved efficiently
using the available software.\\

\emph{\textbf{Theorem 1.} Consider a nonlinear uncertain system of
class $\left({\sum}_{1}\right)$. The output feedback $u=Ky$ robustly
asymptotically stabilizes this system with $\|z\|\leq \mu\|w \|$ and
maximum admissible Lipschitz constant $\gamma^{*}$, if there exist
scalars $\epsilon_{1}>0$, $\epsilon_{1}>0$ and $\alpha>0$ and
matrices $P>0$ and $G$ such that the following LMI optimization
problem has a solution:}
\begin{equation}
\hspace{-6cm} \min (\alpha+\epsilon_{1}) \notag
\end{equation}
\hspace{.5cm} \emph{s.t.}
\begin{align}
&\left[
  \begin{array}{ccccc}
    \Lambda_{1} & I & \Lambda_{2} & 0 & 0 \\
    \star & -\alpha I & 0 & 0 & 0 \\
    \star & \star & -\frac{1}{2}P & P & PM_{1}+GM_{2} \\
    \star & \star & \star & P-2\epsilon_{1}I & 0 \\
    \star & \star & \star & \star & -\epsilon_{2} I \\
  \end{array}
\right]<0\label{LMI8}\\
&\left[
  \begin{array}{ccc}
    -\mu^{2}I & B_{2}^{T}P+D^{T}G^{T} & B_{2}^{T}P+D^{T}G^{T}\\
    \star & -\frac{1}{2}P & 0 \\
    \star & \star & -I\\
  \end{array}
\right]<0\label{LMI9}
\end{align}
\emph{where $\Lambda_{1}=H^{T}H-P+\epsilon_{2} N^{T}N$ and
$\Lambda_{2}=A^{T}P+C^{T}G^{T}$. Once the problem is solved}
\begin{align}
\alpha^{*} &\triangleq \min(\alpha)\notag
\\\epsilon_{1}^{*} &\triangleq \min(\epsilon_{1})\notag
\\\gamma^{*} &\triangleq \max(\gamma)=\frac{1}{\sqrt{\alpha^{*}(1+\epsilon_{1}^{*})}}\notag
\\K&=B_{1}^{T}\overline{K}\notag
\end{align}
\emph{where the matrix $\overline{K}$ is obtained through
\eqref{Kbar1} if and only if} \begin{align}
\textrm{rank}(I_{p}\otimes PB_{1}B_{1}^{T})=\textrm{rank}([
                 \begin{array}{cc}
                   I_{p}\otimes PB_{1}B_{1}^{T} & \textbf{\textrm{vec}}(G) \\
                 \end{array}]).\notag
\end{align}\\
\textbf{Proof:} Substituting $u=Ky$ into \eqref{sigma3_1} yields to
uncertain system
\begin{equation}
\begin{split}
x(k+1)&=(\widetilde{A}+\Delta \widetilde{A}(k))x(k)+\Phi(x,u)+\widetilde{B}w\label{sigma3_2}\\
z(k)&=Hx(k)
\end{split}
\end{equation}
where
\begin{align}
\widetilde{A}&=A+B_{1}KC\notag\\
\Delta \widetilde{A}&=\Delta A+B_{1}K \Delta
C=(M_{1}+B_{1}KM_{2})FN\triangleq \widetilde{M}FN\notag\\
\widetilde{B}&=B_{1}KD+B_{2}\notag.
\end{align}
System \eqref{sigma3_2} is of the form $\sum_{2}$ so Lemma 3
provides a sufficient condition for its robust asymptotic stability.
We have
\begin{align}
P\widetilde{A}&=PA+PB_{1}KC\notag\\
P\widetilde{M}&=PM_{1}+PB_{1}KM_{2}\notag\\
P\widetilde{B}&=PB_{1}KD+PB_{2}\notag.
\end{align}
In order to cast the solution into the form of LMIs, we introduce
the following change of variables:
\begin{align}
PB_{1}K&\triangleq G.\label{G}
\end{align}
Substituting into \eqref{LMI6} and \eqref{LMI7} of lemma 3, the LMIs
\eqref{LMI8} and \eqref{LMI9} are obtained. Without loss of
generality, we assume the feedback gain $K$ to be of the form
\begin{align}
K=B_{1}^{T}\overline{K}\label{gain}
\end{align}
where $\overline{K}\in \mathbb{R}^{n\times p}$ is an unknown matrix
to be found. The algebraic matrix equation \eqref{G} can be solved
for $\overline{K}$ using the Kronecker product as follows
\begin{align}
&PB_{1}B_{1}^{T}\overline{K}=G\Rightarrow (I_{p}\otimes
PB_{1}B_{1}^{T})\textbf{\textrm{vec}}(\overline{K})=\textbf{\textrm{vec}}(G)\label{Kbar1}.
\end{align}
Therefore, there is a solution for
$\textbf{\textrm{vec}}(\overline{K})$ and consequently for $K$ if
and only if
\begin{align}
\textrm{rank}(I_{p}\otimes PB_{1}B_{1}^{T})=\textrm{rank}([
                 \begin{array}{cc}
                   I_{p}\otimes PB_{1}B_{1}^{T} & \textbf{\textrm{vec}}(G) \\
                 \end{array}])\label{Kron_Cond}
\end{align}
which simply means that the subspace spanned by the columns of
$I_{p}\otimes PB_{1}B_{1}^{T}$ must contain
$\textbf{\textrm{vec}}(G)$. Eventually $K$ is obtained from
\eqref{gain}. $\blacksquare$\\

As mentioned in Remark 2, either the admissible Lipschitz constant
or the disturbance attenuation level can be considered as an
optimization variable in Theorem 1. Given this, it may be more
realistic to have a combined performance index. Then,
$\alpha+\epsilon_{1}$ and $\zeta=(\mu^{2})$ can be simultaneously
optimized by convex multiobjective optimization. See
\cite{Abbaszadeh3, abbaszadeh2008robust, abbaszadeh2007robust, abbaszadeh2006robust,
abbaszadeh2010nonlinear, abbaszadeh2008lmi, abbaszadeh2010dynamical, abbaszadeh2010robust,
abbaszadeh2012generalized} for details of the approach and application to
filtering and different classes of nonlinear systems.

Theorem 1 shows that an exact solution for $K$ can be found using
strict LMIs if and only if the condition \eqref{Kron_Cond} is
satisfied. Relaxing this condition, in many cases it is still
possible to find an \emph{approximate stabilizing solution} for $K$
using strict LMIs as discussed in \cite{Abbaszadeh4}. In the
following we derive another formulation for the exact solution of
SOF.

\subsection{Converting BMI into SDP}

In the first part of this section, we showed that an exact solution
for $K$ can be found using strict LMIs if and only if the condition
\eqref{Kron_Cond} is satisfied. Now, we propose another exact
solution by converting the BMI (Bilinear Matrix Inequality) problem
into an SDP (Semidefinite Programming) problem. This adds an
equality constraint to the optimization problem of Theorem 1.\\

\emph{\textbf{Corollary 1.} Consider a nonlinear uncertain system of
class $\left({\sum}_{1}\right)$. The output feedback $u=Ky$ robustly
asymptotically stabilizes this system with $\|z\|\leq \mu\|w \|$ and
maximum admissible Lipschitz constant $\gamma^{*}$, if there exist
scalars $\epsilon_{1}>0$, $\epsilon_{1}>0$ and $\alpha>0$ and
matrices $P>0$, $Q$ and $G$ such that the following SDP problem has
a solution:}
\begin{equation}
\hspace{-6cm} \min (\alpha+\epsilon_{1}) \notag
\end{equation}
\hspace{.3cm} \emph{s.t.}
\begin{align}
&\ PB_{1}=B_{1}Q\\
&\Pi_{1}\triangleq\left[
    \begin{array}{cc}
      I & I-Q \\
      \star & I \\
    \end{array}
  \right]>0\label{LMI10}\\
&\Pi_{2}\triangleq\left[
  \begin{array}{ccc}
    -\mu^{2}I & \Lambda_{4} & \Lambda_{4}\\
    \star & -\frac{1}{2}P & 0 \\
    \star & \star & -I\\
  \end{array}
\right]<0\\
&\left[
  \begin{array}{ccccc}
    \Lambda_{1} & I & \Lambda_{3} & 0 & 0 \\
    \star & -\alpha I & 0 & 0 & 0 \\
    \star & \star & -\frac{1}{2}P & P & PM_{1}+B_{1}GM_{2} \\
    \star & \star & \star & P-2\epsilon_{1}I & 0 \\
    \star & \star & \star & \star & -\epsilon_{2} I \\
  \end{array}
\right]<0
\end{align}
\emph{where $\Lambda_{1}$ is as in Theorem 1,
$\Lambda_{3}=A^{T}P+C^{T}G^{T}B_{1}^{T}$ and
$\Lambda_{4}=B_{2}^{T}P+D^{T}G^{T}B_{1}^{T}$. Once the problem is
solved}
\begin{align}
\alpha^{*} &\triangleq \min(\alpha)\notag
\\\epsilon_{1}^{*} &\triangleq \min(\epsilon_{1})\notag
\\\gamma^{*} &\triangleq \max(\gamma)=\frac{1}{\sqrt{\alpha^{*}(1+\epsilon_{1}^{*})}}\notag
\\K&=Q^{-1}G\notag
\end{align}\\
\textbf{Proof:} Using the change of variable $G=QK$, we have
$PB_{1}K=B_{1}QK\triangleq B_{1}G$. Having $G$, there exists a
unique exact solution for $K$ if and only if $Q$ is nonsingular. The
LMI \eqref{LMI10} guarantees the nonsingularity of $Q$ \cite[p.
312]{Horn1}. The rest of the proof is the same as the proof of
Theorem 1. $\blacksquare$\\

Note that $Q_{m\times m}$ only needs to be nonsingular and is not
necessarily positive definite or even symmetric. The SDP problem of
Corollary 2 can be
solved by freely available packages such as YALMIP \cite{YALMIP}. \\

\emph{\textbf{Remark 4.}} Suppose that instead of \eqref{output1},
the output map in $\left({\sum}_{3}\right)$ is as
\begin{align}
y(k)=(C+\Delta C(k))x(k)+D_{1}u+D_{2}w.\notag
\end{align}
Defining the fictitious output $\overline{y}=y-D_{1}u=(C+\Delta
C(k))x(k)+D_{2}w$, we first find the static output feedback
$u=K\overline{y}$. Then we have $u=(I+KD_{1})^{-1}Ky$ provided that
the inverse exists.

\section{Robustness Against Nonlinear Uncertainty}
As mentioned earlier, the maximization of Lipschitz constant makes
the proposed observer robust against some Lipschitz nonlinear
uncertainty. In this section this robustness feature is studied and
both norm-vise and element-wise bounds on the nonlinear uncertainty
are derived. The norm-wise analysis provides an upper bound on the
Lipschitz constant of the nonlinear uncertainty and the norm of the
Jacobian matrix of the corresponding nonlinear function.
Furthermore, we will find upper and lower bounds on the elements of
the \emph{matrix-type Lipschitz constant} of the nonlinear
uncertainty through a novel element-wise analysis.

\subsection{Norm-Wise Robustness}
Assume a nonlinear uncertainty as follows
\begin{align}
\Phi_{\Delta}(x,u)&=\Phi(x,u)+\Delta\Phi(x,u)
\\x(k+1)&=(A+ \Delta A)x(k) + \Phi_{\Delta}(x,u)
\label{uncer3}
\end{align}
where $\Phi_{\Delta}$ is the uncertain nonlinear function and
$\Delta\Phi$ is the unknown nonlinear uncertainty. Suppose that
\begin{eqnarray}
\|\Delta\Phi(x_{1},u)-\Delta\Phi(x_{2},u)\|\leqslant\Delta\gamma\|x_{1}-x_{2}\|.\\
\notag
\end{eqnarray}

\emph{\textbf{Proposition 1.}} {\emph{Suppose that the actual
Lipschitz constant of the system is $\gamma$ and the maximum
admissible Lipschitz constant achieved by Corollary 1 (Theorem 1),
is $\gamma^{*}$. Then, the observer designed based on Corollary 1
(Theorem 1), can tolerate any additive Lipschitz nonlinear
uncertainty with Lipschitz constant less than or
equal $\gamma^{*}-\gamma$}}.\\

\textbf{Proof:} Based on Schwartz inequality, we have
\begin{eqnarray}
\begin{split}
\|\Phi_{\Delta}(x_{1},u)&-\Phi_{\Delta}(x_{2},u)\|\leq
\|\Phi(x_{1},u)-\Phi(x_{2},u)\|\\&+\|\Delta\Phi(x_{1},u)-\Delta\Phi(x_{2},u)\|\leq
\gamma\|x_{1}-x_{2}\|\\
&+\Delta\gamma\|x_{1}-x_{2}\|.
\end{split}\notag
\end{eqnarray}
According to the Corollary 1 (Theorem 1), $\Phi_{\Delta}(x,u)$ can
be any Lipschitz nonlinear function with Lipschitz constant less
than or equal to $\gamma^{*}$,
\begin{equation}
\|\Phi_{\Delta}(x_{1},u)-\Phi_{\Delta}(x_{2},u)\|\leq\gamma^{*}\|x_{1}-x_{2}\|,\notag
\end{equation}
so, there must be
\begin{eqnarray}
\gamma+\Delta\gamma\leq\gamma^{*}\rightarrow\Delta\gamma\leq\gamma^{*}-\gamma.\
\ \ \blacksquare \notag
\end{eqnarray}\\
In addition, we know that for any continuously differentiable
function $\Delta\Phi$, $\forall \ x, x_{1},x_{2} \in \mathcal{D}$
\begin{eqnarray}
\|\Delta\Phi(x_{1},u)-\Delta\Phi(x_{2},u)\|\leqslant\|\frac{\partial\Delta\Phi}{\partial
x}(x_{1}-x_{2})\|,\notag
\end{eqnarray}
where $\frac{\partial\Delta\Phi}{\partial x}$ is the Jacobian matrix
\cite{Marquez}. So $\Delta\Phi(x,u)$ can be any additive uncertainty
with $\|\frac{\partial\Delta\Phi}{\partial x}\| \leq
\gamma^{*}-\gamma$. In the following we will find approperate bound
on the entris of the Jacobian matrix.

\subsection{Element-Wise Robustness}
Assume that there exists a matrix $\Gamma\in\mathbb{R}^{n\times n}$
such that
\begin{equation}
\|\Phi(x_{1},u)-\Phi(x_{2},u)\|\leqslant\|\Gamma(x_{1}-x_{2})\|.\label{Gamma}
\end{equation}
$\Gamma$ can be considered as a \emph{matrix-type Lipschitz
constant}. Suppose that the nonlinear uncertainty is as in
\eqref{uncer3} and
\begin{equation}
\|\Phi_{\Delta}(x_{1},u)-\Phi_{\Delta}(x_{2},u)\|\leqslant\|\Gamma_{\Delta}(x_{1}-x_{2})\|.\label{Gamma-Delta}
\end{equation}
Assuming
\begin{equation}
\|\Delta\Phi(x_{1},u)-\Delta\Phi(x_{2},u)\|\leqslant\|\Delta\Gamma(x_{1}-x_{2})\|,\label{uncer4}
\end{equation}
based the proposition 1, $\Delta\Gamma$ can be any matrix with
$\|\Delta\Gamma\|\leq \gamma^{*}-\|\Gamma\|$. Now, we look at the
problem from a different angle. It is clear that
$\Gamma_{\Delta}=[{\gamma_{\Delta}}_{ij}]_{n}$ is a perturbed
version of $\Gamma$ due to $\Delta \Phi(x,u)$. The question is that
how much perturbation can be tolerated on the elements of $\Gamma$
without loosing the observer features stated in
Corollary 1 (Theorem 1). In the following we will find appropriate bound
on the entries of the Jacobian matrix., but how about the entries of $\Delta\Gamma$?
This is important in the sense that in gives us an insight about the
amount of uncertainty that can be tolerated in different directions
of the nonlinear function. Here, we propose an approach to optimize
the elements of $\Gamma$ and provide specific upper and lower
bounds on tolerable perturbations.\\

\emph{\textbf{Corollary 2.} Consider Lipschitz nonlinear system
$\left(\sum_{1} \right)$ satisfying \eqref{Gamma}, along with the
control law $u=Ky$. The closed loop system is (globally)
asymptotically stable with the matrix-type Lipschitz constant
$\Gamma^{*}=[\gamma^{*}_{ij}]_{n}$ with maximized admissible
elements and $\mathfrak{L}_{2}(w \rightarrow z)$ gain, $\mu$, if
there exist fixed scalars $\mu>0$ and $c_{ij}>0 \ \forall \ 1\leq
i,j\leq n$, scalars $\omega>0$, $\epsilon_{1}>0$ and
$\epsilon_{2}>0$ and matrices $\mathcal{A}=[\alpha_{ij}]_{n}\succ
0$, $P_{1}>0$, $P_{2}>0$ and $G$, such that the following LMI
optimization problem has a solution.}\\
\begin{equation}
\hspace{-4cm} \min \ (\epsilon_{1}-\omega) \notag
\end{equation}
\hspace{0.3cm} \emph{s.t.}
\begin{align}
&c_{ij}\alpha_{ij}>\omega \ \ \ \forall \ 1\leq i,j\leq n\notag\\
&PB_{1}=B_{1}Q\notag\\
&\Pi_{1}>0\notag\\
&\Pi_{2}<0 \notag\\
&\left[
  \begin{array}{ccccc}
    \Lambda_{1} & \mathcal{A} & \Lambda_{3} & 0 & 0 \\
    \star & - I & 0 & 0 & 0 \\
    \star & \star & -\frac{1}{2}P & P & PM_{1}+B_{1}GM_{2} \\
    \star & \star & \star & P-2\epsilon_{1}I & 0 \\
    \star & \star & \star & \star & -\epsilon_{2} I \\
  \end{array}
\right]<0\notag
\end{align}
\emph{where $\Pi_{1}$, $\Pi_{2}$, $\Lambda_{1}$ and $\Lambda_{3}$
are as in Corollary 1. Once the problem is solved}
\begin{eqnarray}
L&=&P_{1}^{-1}G\notag\\
\alpha^{*}_{ij} &\triangleq& \max(\alpha_{ij})\notag\\
\mathcal{A}^{*}&\triangleq& [\alpha^{*}_{ij}]_{n}\notag\\
\epsilon_{1}^{*} &\triangleq& \min(\epsilon_{1})\notag \\
\Gamma^{*}&\triangleq&
\frac{1}{\sqrt{(1+\epsilon_{1}^{*})}}\mathcal{A}^{*}\notag
\end{eqnarray}\\
\textbf{Proof:} The proof is similar to the proof of Corollary 1,
replacing $\gamma I$ with $\Gamma$ and using the change of variables
$(1+\epsilon_{1})\Gamma^{T}\Gamma=\mathcal{A}^{T}\mathcal{A}$ . \ \ \ $\blacksquare$\\

\textbf{Remark 3.} By appropriate selection of the weights
$c_{i,j}$, it is possible to put more emphasis on the directions in
which the tolerance against nonlinear uncertainty is more important.
To this goal, one can take advantage of the knowledge
about the structure of the nonlinear function $\Phi(x,u)$.\\

Note that the robustness results of this section can be similarly
applied to Theorem 1, as well. According to the norm-vise analysis,
it is clear that $\Delta\Gamma$ in \eqref{uncer4} can be any matrix
with $\|\Delta\Gamma\|\leq \|\Gamma^{*}\|-\|\Gamma\|$. We now
proceed by deriving bounds on the elements of $\Gamma_{\Delta}$. To
prove our result we recall the following lemma from
\cite{Abbaszadeh3}.\\

\emph{\textbf{Lemma 4. \cite{Abbaszadeh3}} For any $S=[s_{i,j}]_{n}$
and $T=[t_{i,j}]_{n}$, if $|S|\preceq T$, then $SS^{T}\leq
TT^{T}\circ
nI$}.\\


Now we are ready to state the element-wise robustness result.\\ Assume
additive uncertainty in the form of \eqref{uncer3}, where
\begin{equation}
\|\Phi_{\Delta}(x_{1},u)-\Phi_{\Delta}(x_{2},u)\|\leqslant\|\Gamma_{\Delta}(x_{1}-x_{2})\|.\label{Gamma-Delta}
\end{equation}
It is clear that $\Gamma_{\Delta}=[{\gamma_{\Delta}}_{i,j}]_{n}$ is
a perturbed version of $\Gamma$.\\

\emph{\textbf{Proposition 2.}} \emph{Suppose that the actual
matrix-type Lipschitz constant of the system is $\Gamma$ and the
maximized admissible matrix-type Lipschitz constant achieved by
Corollary 1 (Theorem 1), is $\Gamma^{*}$. Then, $\Delta\Phi$ can be
any additive nonlinear uncertainty such that
$|\Gamma_{\Delta}|\preceq
n^{-\frac{1}{2}} \Gamma^{*}$.}\\

\textbf{Proof:} According to the Proposition 2, it suffices to show
that $\sigma_{max}(\Gamma_{\Delta})\leq\sigma_{max}(\Gamma^{*})$.
Using Lemma 4, we have
\begin{equation}
\begin{split}
\sigma_{max}^{2}(\Gamma_{\Delta})&=\lambda_{max}(\Gamma_{\Delta}\Gamma_{\Delta}^{T})\\
&\leq\lambda_{max}(n^{-1}\Gamma^{*}{\Gamma^{*}}^{T}\circ nI)\\
&=\sigma_{max}(\Gamma^{*}{\Gamma^{*}}^{T}\circ I)\\
&\leq
\sigma_{max}(\Gamma^{*}{\Gamma^{*}}^{T})=\sigma_{max}^{2}(\Gamma^{*})
\end{split}\notag
\end{equation}
The first inequality follows from lemma 4 and the symmetry of
$\Gamma_{\Delta}\Gamma_{\Delta}^{T}$ and
diag(diag($\Gamma^{*}{\Gamma^{*}}^{T}))$, \cite[p. 200]{Horn1}. The
last inequality is due to the fact that the spectral norm is
submultiplicative with respect to the Hadamard product \cite[Ch.
5]{Horn2}. Since the singular values are nonnegative, we can
conclude that
$\sigma_{max}(\Gamma_{\Delta})\leq\sigma_{max}(\Gamma^{*})$. \ \ \
$\blacksquare$\\

Therefore, denoting the element of $\Gamma_{\Delta}$ as
${\gamma_{\Delta}}_{i.j}=\gamma_{i,j}+\delta_{i,j}$, the following
bound on the element-vise perturbations is obtained
\begin{equation}
-n^{-\frac{1}{2}}\gamma^{*}_{i,j}-\gamma_{i,j} \leq \delta_{i,j}
\leq n^{-\frac{1}{2}}\gamma^{*}_{i,j}-\gamma_{i,j}.
\end{equation}
In addition, $\Delta\Phi(x,u)$ can be any continuously
differentiable additive uncertainty which makes
$|\frac{\partial\Phi_{\Delta}}{\partial x}|\preceq n^{-\frac{1}{2}}
\Gamma^{*}$.

It is worth mentioning that the results of Lemma 4 and Proposition 2
have intrinsic importance from the matrix analysis point of view
regardless of our specific application in the robustness analysis.

\section{Numerical Example}

\begin{figure}[!h]
  \centering
  \includegraphics[width=4.5in]{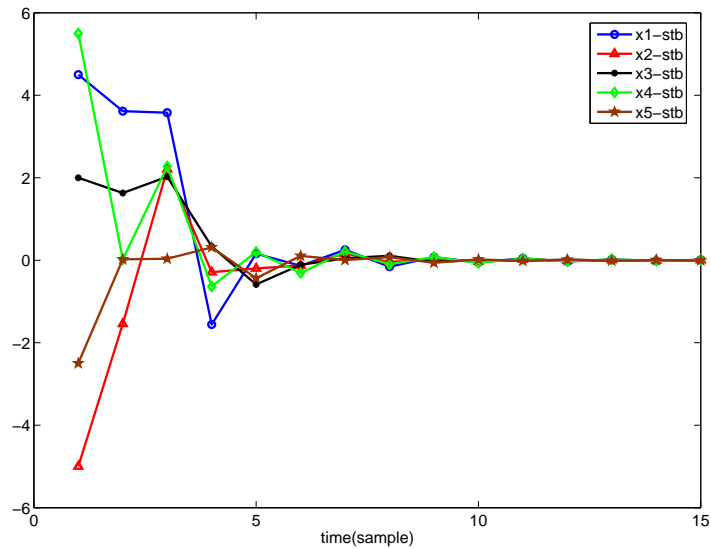}\\
  \caption{The stabilized state trajectories}\label{Fig4}
\end{figure}

I this example, we design an static output feedback controller for
an uncertain nonlinear system. Consider a system of class
$(\sum_{1})$ where, {\small \begin{align}
 &A=\left[
    \begin{array}{ccccc}
    0.5000 &  -0.5975 &   0.3735 &   0.0457 &   0.3575\\
    0.2500 &   0.3000 &   0.4017 &   0.1114 &   0.0227\\
    0.4880 &   0.1384 &   0.2500 &   0.7500 &   0.7500\\
    0.3838 &   0.0974 &   0.5000 &   0.2500 &   0.5000\\
    0.0347 &   0.1865 &  -0.2500 &   0.5000 &   0.2500
    \end{array}
  \right], \notag\\
&\Phi(x,u)=\left[
                              \begin{array}{c}
                                0.1\sin(x_{3}) \\
                                0.2\sin(x_{4}) \\
                                0.3\sin(x_{1}) \\
                                0 \\
                                0.1\sin(x_{2})
                              \end{array}
                            \right], C=\left[
                                                        \begin{array}{cc}
                                                               0.5 &    0\\
                                                               0.2 &  0.2\\
                                                               0   &  0.1\\
                                                               0   &  0.3\\
                                                               0.3 &    0\\
                                                        \end{array}
                                                   \right], B_{2}=\left[
                                \begin{array}{c}
                                  1 \\
                                  1 \\
                                  1 \\
                                  1 \\
                                  1 \\
                                \end{array}
                              \right], \notag\\
                              &B_{1}=\left[
              \begin{array}{ccc}
                 0.7  &  0.8  &    0\\
                 0.4  &  0.9  &  0.9\\
                 0.9  &  0.9  &  0.2\\
                 0.9  &  0.6  &  0.7\\
                   0  &  0.5  &  0.3\\
              \end{array}
         \right], M_{1}=\left[
                                                        \begin{array}{cc}
                                                               0.1  &    0\\
                                                               0.1  &  0.1\\
                                                               0.1  &  0.1\\
                                                                 0  &  0.1\\
                                                                 0  &  0.2\\
                                                        \end{array}
                                                   \right], D=\left[
                                            \begin{array}{c}
                                                   0.2 \\
                                                   0.2 \\
                                            \end{array}
                                        \right], \notag\\
&M_{2}=\left[
        \begin{array}{cc}
            0 & 0.1 \\
          0.1 & 0.2 \\
        \end{array}
      \right], N=\left[
                         \begin{array}{ccccc}
                           0.3 & 0.15 & 0.1 & 0 & 0.2 \\
                           0.1 & 0.2  & 0.1 & 0.2 & 0 \\
                         \end{array}
                       \right].\notag
\end{align}}
The system is globally Lipschitz with Lipschitz constant $0.3$. The
nonlinear system is unstable. The matrix $A$ is unstable, which
means the linear part of the systems is itself unstable. The matrix
$A$ is singular, so the approach of \cite{Garcia} is not applicable
even to the nominal linear part. The submatrix of $A$ obtained by
omitting the first $p(=2)$ rows and columns in singular, too.
Therefore, we can not use the results of \cite{Bara1}, neither. Now,
we use Corollary 1 with $H=0.15I_{5}$ and $\mu=2.5$, to design $K$.
Using YALMIP, we solve the proposed SDP problem and we get:
\begin{align*}
\epsilon_{1}^{*}&=0.2076, \ \ \alpha^{*}=0.3013, \ \ \gamma^{*}=1.6584\\
K&=\left[
     \begin{array}{cc}
        -0.7026 &  -0.8334\\
         0.4825 &  -1.8664\\
         0.3292 &  -1.1758\\
     \end{array}
   \right].
\end{align*}
Figure \ref{Fig4} shows the state trajectories of the stabilized
system.


\section{Conclusions}

A new LMI optimization approach to the robust static output feedback
stabilization for nonlinear discrete-time uncertain is systems with
$H_{\infty}$ performance is proposed. The considered class of
nonlinear systems contains norm-bounded time-varying model
uncertainties as well as additive Lipschitz nonlinear model
uncertainties. Explicit bounds on the tolerable uncertainty were
derived via norm-wise and element-wise robustness analysis.


\bibliographystyle{IEEEtran}
\bibliography{Candidacy_References}


\end{document}